\documentclass[12pt]{article} 

\usepackage{amssymb} 
\usepackage{amsmath}
\usepackage{amsfonts}

\newcommand{\R}{\mathbb{R}}
\newcommand{\C}{\mathbb{C}}
\newcommand{\Z}{\mathbb{Z}}

\newcommand{\be}{\begin{equation}}
\newcommand{\bea}{\begin{eqnarray}}
\newcommand{\eea}{\end{eqnarray}}
\newcommand{\nn}{\nonumber}
\newcommand{\kt}{\rangle}
\newcommand{\br}{\langle}

\newcommand{\ed}{\end{document}}
\newcommand{\bbr}{\br\!\br}
\newcommand{\kkt}{\kt\!\kt}

\begin{document}

\title{Pseudo-Hermiticity for a Class of Nondiagonalizable Hamiltonians}
\author{Ali Mostafazadeh\thanks{E-mail address: amostafazadeh@ku.edu.tr}\\ \\
Department of Mathematics, Ko\c{c} University,\\
Rumelifeneri Yolu, 80910 Sariyer, Istanbul, Turkey}
\date{ }
\maketitle

\begin{abstract}
We give two characterization theorems for pseudo-Hermitian (possibly nondiagonalizable) Hamiltonians with a discrete spectrum that admit a block-diagonalization with finite-dimensional diagonal blocks. In particular, we prove that for such an operator $H$ the following statements are equivalent. 1.~$H$ is pseudo-Hermitian; 2.~The spectrum of $H$ consists of real and/or complex-conjugate pairs of eigenvalues and the geometric multiplicity and the dimension of the diagonal blocks for the complex-conjugate eigenvalues are identical; 3.~$H$ is Hermitian with respect to a positive-semidefinite inner product. We further discuss the relevance of our findings for the merging of a complex-conjugate pair of eigenvalues of diagonalizable pseudo-Hermitian Hamiltonians in general, and the $PT$-symmetric Hamiltonians and the effective Hamiltonian for a certain closed FRW minisuperspace quantum cosmological model in particular.
\end{abstract}

\baselineskip=24pt

\section{Introduction}

In Refs.~\cite{p1,p2,p3,p4,p6} we developed the notion of a pseudo-Hermitian operator and investigated its various consequences in particular in connection with $PT$-symmetric quantum systems \cite{pt} and two-component formulation of the FRW minisuperspace quantum cosmology \cite{jmp-98}. Since the announcement of the results of \cite{p1} several authors have explored the implications of pseudo-Hermiticity, \cite{others}. The main results reported in Refs.~\cite{p1,p2,p3,p4,p6} were however based on the assumption that the Hamiltonian of the system is diagonalizable and has a discrete spectrum. As demonstrated in \cite{p6}, the latter condition can be easily relaxed. Moreover in Ref.~\cite{p5} we showed, without making any assumption about the diagonalizability of the Hamiltonian or discreteness of its spectrum, that the results of Refs.~\cite{p1} and \cite{p3} generalized to the class of all $PT$-symmetric standard Hamiltonians having $\R$ as their configuration space. This suggests that these results may be valid under more general conditions.
The purpose of the present article is to generalize the results of \cite{p1} to the class of possibly nondiagonalizable Hamiltonians that admit a block-diagonalization with finite-dimensional diagonal blocks. This, in particular, includes all the matrix Hamiltonians. It is also relevant to the accidental loss of diagonalizability due to the pseudo-Hermiticity-preserving variations of diagonalizable pseudo-Hermitian Hamiltonians that lead to the merging of complex-conjugate pairs of eigenvalues.

The organization of the article is as follows. In Section~2, we discuss basic properties of the class of the Hamiltonians admitting a block-diagonalization with finite-dimensional diagonal blocks. In Section~3, we present two characterization theorems for pseudo-Hermitian Hamiltonians belonging to this class. In Section~4, we study general $2\times 2$ matrix Hamiltonians. In Section~5, we discuss an application of our results in quantum cosmology. Finally, in Section~6, we present our concluding remarks.

\section{Block-diagonalizable Hamiltonians with Finite-\ Dimensional Diagonal Blocks}

Consider a linear operator $H:{\cal H}\to{\cal H}$ acting in a (separable) Hilbert space ${\cal H}$
and having a discrete spectrum. Suppose that for every eigenvalue $E_n$, there are positive integers 
$g_n, p_n\in\Z^+$ such that for all $\ell\in\Z^+$,
	\be
	d_{n,\ell}:={\rm dim}[{\rm ker}(H-E_n 1)^\ell] = g_n~~~~\mbox{if and only if}~~~~
	\ell\geq p_n.
	\label{eq1}
	\end{equation}
This in particular means that
	\be
	d_{n,1}\leq d_{n,2}\leq\cdots\leq d_{n,p_n-1}\leq d_{n,p_n}=g_n.
	\label{eq2}
	\end{equation}
The integer $d_{n,1}$ is just the degree of degeneracy or the geometric multiplicity of $E_n$. In what follows, we shall use the abbreviated notation $d_n$ for $d_{n,1}$ and denote the degeneracy labels $1,2\cdots,d_n$ by the letters from the beginning of the Latin alphabet.

The integer $g_n$ is called the algebraic multiplicity of $E_n$. The condition (\ref{eq1}) means that all the eigenvalues of $H$ have finite algebraic multiplicity. Throughout this paper we shall assume that this condition is satisfied and that there is a basis of the Hilbert space in which $H$ is block-diagonal with diagonal blocks being finite-dimensional. In this case, we can always find a basis in which the diagonal blocks have the canonical Jordan form \cite{prasolov}, i.e., there is an invertible operator 
$A:{\cal H}\to{\cal H}$ and an orthonormal basis $\{|n,a,i\kt\}$ with $n$ being the spectral label, $a\in\{1,2,\cdots,d_n\}$, $i\in\{1,2,\cdots,p_{n,a}\}$, and $p_{n,a}\in\Z^+$, such that
	\be
	A^{-1}HA=H_b:=\sum_n\sum_{a=1}^{d_n}
	\left(E_n\sum_{i=1}^{p_{n,a}}|n,a,i\kt\br n,a,i|+
	\sum_{i=1}^{p_{n,a}-1}|n,a,i\kt\br n,a,i+1|\right).
	\label{H_b}
	\end{equation}
Alternatively, letting 
	\be
	|\psi_n,a,i\kt:=A|n,a,i\kt,~~~~|\phi_n,a,i\kt:=A^{-1\dagger}|n,a,i\kt,
	\label{pp=}
	\end{equation}
we have 
	\bea
	&&\br\psi_n,a,i|\phi_m,b,j\kt=\delta_{mn}\delta_{ab}\delta_{ij},
	~~~~\sum_i\sum_{a=1}^{d_n}\sum_{i=1}^{p_{n,a}}|\psi_n,a,i\kt\br\phi_n,a,i|=1,
	\label{bior}\\
	&&H=AH_bA^{-1}=
	\sum_n\sum_{a=1}^{d_n}\left(E_n\sum_{i=1}^{p_{n,a}}|\psi_n,a,i\kt\br\phi_n,a,i|+
	\sum_{i=1}^{p_{n,a}-1}|\psi_n,a,i\kt\br\phi_n,a,i+1|\right).
	\label{H=H_b}
	\eea
Note that according to Eqs.~(\ref{bior}) and (\ref{H=H_b}), $\{|\psi_n,a,i\kt,|\phi_n,a,i\kt\}$ is a complete biorthonormal system for the Hilbert space and 
	\bea
	H|\psi_n,a,1\kt&=&E_n|\psi_n,a,1\kt\,,
	\label{e1}\\
	H^\dagger|\phi_n,a,p_{n,a}\kt&=&E_n^*|\phi_n,a,p_{n,a}\kt\,.
	\label{e2}
	\eea
Hence $|\psi_n,a,1\kt$ are the eigenvectors of $H$ and $|\phi_n,a,p_{n,a}\kt$ are the eigenvectors of $H^\dagger$. 

The numbers $p_{n,a}$ represent the dimension of the Jordan block associated with the spectral label $n$ and the degeneracy label $a$. We shall refer to them as the Jordan dimensions. For a given eigenvalue $E_n$, the number of the corresponding Jordan blocks (which is equal to the geometric multiplicity of $E_n$) and the Jordan dimensions are uniquely determined by the integers $d_{n,\ell}$ of (\ref{eq1}) up to the permutations of the degeneracy labels, \cite{prasolov}. Note also that the algebraic multiplicity is the sum of the Jordan dimensions, $g_n:=\sum_{a=1}^{d_n}p_{n,a}$.

\section{Consequences of Pseudo-Hermiticity}

\begin{itemize}
\item[] {\bf Theorem~1:} Let $H:{\cal H}\to{\cal H}$ be a linear operator acting in a (separable) Hilbert space ${\cal H}$. Suppose that the spectrum of $H$ is discrete, its eigenvalues have finite  algebraic multiplicity, and that (\ref{H=H_b}) holds. Then, $H$ is pseudo-Hermitian if and only if the eigenvalues of $H$ are either real or come in complex-conjugate pairs and the geometric multiplicity and the Jordan dimensions of the complex-conjugate eigenvalues coincide.
\item[] {\bf Proof:} Suppose that $H$ is pseudo-Hermitian.  Then, by definition \cite{p1}, there is a Hermitian automorphism (linear bijection mapping ${\cal H}$ onto ${\cal H}$) $\eta:{\cal H}\to{\cal H}$ such that $H^\dagger=\eta H\eta^{-1}$. Now let $E_n$ be an arbitrary element of the spectrum of $H$. Then, by virtue of Eqs.~(\ref{e1}) and (\ref{e2}), for each $a\in\{1,2,\cdots,d_n\}$, $|\psi_n,a,1\kt$ is an eigenvector of $H$ with eigenvalue $E_n$ and $|\phi_n,a,p_{n,a}\kt$ is an eigenvector of $H^\dagger$ with eigenvalue $E_n^*$. This in turn implies 
$H\eta^{-1}|\phi_n,a,p_{n,a}\kt=\eta^{-1}H^\dagger|\phi_n,a,p_{n,a}\kt=
E_n^*\eta^{-1}|\phi_n,a,p_{n,a}\kt$. As $\eta^{-1}$ is an invertible operator, $\eta^{-1}|\phi_n,a,p_{n,a}\kt\neq 0$. Hence $E_n^*$ also belongs to the spectrum of $H$. Next, note that because the eigenvalues of $H$ and consequently $H^\dagger$ have 
finite algebraic multiplicity, for every $\ell\in\Z^+$, ${\rm kernel}(H-E_n)^\ell$, 
${\rm kernel}(H-E_n^*)^\ell$, and ${\rm kernel}(H^\dagger-E_n^*)^\ell$ are finite-dimensional 
subspaces of ${\cal H}$. Clearly, as a result of (\ref{H=H_b}), $H$ and $H^\dagger$ have essentially
the same Jordan block-diagonalization. In particular, the geometric multiplicity and the Jordan dimensions of $E_n^*$ as an eigenvalue of $H^\dagger$ is the same as the geometric multiplicity and the Jordan dimensions of $E_n$ as an eigenvalue of $H$. This implies that ${\rm kernel}(H^\dagger-E_n^*)^\ell$ and ${\rm kernel}(H-E_n)^\ell$ have the same dimension. Thus they are isomorphic. Furthermore, using the fact that $\eta$ is an automorphism, ${\rm kernel}(H^\dagger-E_n^*)^\ell$ is also isomorphic to
	\[ {\rm kernel} [\eta^{-1}(H^\dagger-E_n^*)^\ell\eta]=
	{\rm kernel}(\eta^{-1}H^\dagger\eta-E_n^*)^\ell=
	{\rm kernel}(H-E_n^*)^\ell.\]
Therefore, for every $\ell\in\Z^+$, ${\rm kernel}(H-E_n)^\ell$ and ${\rm kernel}(H-E_n^*)^\ell$ are isomorphic and consequently have the same dimension. This in turns implies that the number of the Jordan blocks associated with $E$ and their dimensions are identical with those of $E_n^*$, i.e., $E_n$ and $E_n^*$ have the same geometric multiplicity, and up to permutations of the degeneracy labels they have identical Jordan dimensions as well. Conversely, suppose that the eigenvalues of $H$ are either real or come in complex-conjugate pairs and the geometric multiplicity $d_n$ and the Jordan dimensions $p_{n,a}$ of the complex conjugate pairs of eigenvalues are identical. We shall set $n=\nu_0,\nu,\nu-$ depending on whether imaginary part of $E_n$ is zero, positive, or negative. Then $E_{\nu-}=E_\nu^*$, $d_{\nu-}=d_\nu$, for all $a\in\{1,2,\cdots,d_n\}$, $p_{\nu-,a}=p_{\nu,a}$, and Eq.~(\ref{H=H_b}) takes the form
	\bea
	H&=&\sum_{\nu_0}\sum_{a=1}^{d_{\nu_0}}
	\left(E_{\nu_0}\sum_{i=1}^{p_{\nu_0,a}}|\psi_{\nu_0},a,i\kt\br\phi_{\nu_0},a,i|+
	\sum_{i=1}^{p_{\nu_0,a}-1}|\psi_{\nu_0},a,i\kt\br\phi_{\nu_0},a,i+1|\right)+\nn\\
	&&\sum_{\nu}\sum_{a=1}^{d_{\nu}}\left[\sum_{i=1}^{p_{\nu,a}}
	\left(E_{\nu} |\psi_{\nu},a,i\kt\br\phi_{\nu},a,i|+
	E_\nu^*|\psi_{\nu-},a,i\kt\br\phi_{\nu-},a,i|\right)+\right.\nn\\
	&&\left.\sum_{i=1}^{p_{\nu,a}-1}
	\left(|\psi_{\nu},a,i\kt\br\phi_{\nu},a,i+1|+|\psi_{\nu-},a,i\kt\br\phi_{\nu-},a,i+1|\right)
	\right].
	\label{e3}
	\eea
Next, let
	\bea
	\eta(x,\xi)&:=&\sum_{\nu_0}\sum_{a=1}^{d_{\nu_0}}
	\sum_{i=1}^{p_{\nu_0,a}}\sum_{j=p_{\nu_0,a}+1-i}^{p_{\nu_0,a}}
	x_{\nu_0,a,i+j}|\phi_{\nu_0},a,i\kt\br\phi_{\nu_0},a,j|+\nn\\
	&&\sum_{\nu}\sum_{a=1}^{d_{\nu}}
	\sum_{i=1}^{p_{\nu,a}}\sum_{j=p_{\nu,a}+1-i}^{p_{\nu,a}}
	(\xi_{\nu,a,i+j}|\phi_{\nu},a,i\kt\br\phi_{\nu-},a,j|+
	\xi_{\nu,a,i+j}^*|\phi_{\nu-},a,j\kt\br\phi_{\nu},a,i|),\nn\\
	&&
	\label{e4}
	\eea
where $x_{\nu_0,a,k}\in\R$, $\xi_{\nu,a,k}\in\C$, 
	\be
	x_{\nu_0,a,p_{\nu_0,a}+1}\neq 0\neq \xi_{\nu,a,p_{\nu,a}+1},
	\label{e5}
	\end{equation}
and $x$ and $\xi$ respectively stand for the sequences $\{x_{\nu_0,a,k}\}$ and $\{\xi_{\nu,a,k}\}$. 
It is not difficult to check that, for all $n=\nu_0,\nu,\nu-$, $m=\mu_0,\mu,\mu-$, and the corresponding degeneracy labels $a,b$ and Jordan block labels $i,j$, 
	\bea
	&&\br\psi_{\nu_0},a,i|\eta(x,\xi)|\psi_{\mu_0},b,j\kt=\left\{\begin{array}{cc}
	\delta_{\nu_0,\mu_0}\delta_{ab}x_{\nu_0,a,i+j}&{\rm for}~~~ i+j>p_{\nu_0,a}\\
	0 &{\rm otherwise}\end{array}\right.,
	\label{e5.1}\\
	&&\br\psi_{\nu},a,i|\eta(x,\xi)|\psi_{\mu-},b,j\kt=
	\br\psi_{\mu-},a,i|\eta(x,\xi)|\psi_{\nu},b,j\kt^*=
	\left\{\begin{array}{cc}
	\delta_{\nu,\mu}\delta_{ab}\xi_{\nu,a,i+j}&{\rm for}~~~ i+j>p_{\nu,a}\\
	0 &{\rm otherwise}\end{array}\right.,\nn\\
	&&\label{e5.2}
	\eea
and that the other matrix elements of $\eta:=\eta(x,\xi)$, in the basis $\{|\psi_n,a,j\kt\}$, vanish. In view of Eqs.~(\ref{e5.1}), (\ref{e5.2}) and (\ref{e5}), $\eta$ is a Hermitian automorphism. Furthermore, using Eqs.~(\ref{bior}) and (\ref{e3}) -- (\ref{e5}), one can check that it satisfies
$\eta H=H^\dagger\eta$. Hence, $H^\sharp:=\eta^{-1}H^\dagger \eta=H$, and $H$ is $\eta$-pseudo-Hermitian.~~$\square$
\end{itemize}
An immediate consequence of this theorem is
\begin{itemize}
\item[] {\bf Corollary~1:} Let $H$ be as in Theorem~1. Then the pseudo-Hermiticity of $H$ is a necessary condition for the reality of its spectrum.
\end{itemize}

Note that (\ref{e4}) is not the most general expression for an $\eta$ with respect to which $H$ is $\eta$-pseudo-Hermitian. One can obtain more general expressions by performing appropriate 
basis transformations.\footnote{These are the transformations that mix the basis vectors with
different degeneracy labels $a$ but identical spectral label $n$ and the Jordan dimension $p_{n,a}$.} Similarly to the diagonalizable case \cite{p4}, one can also perform a change of basis to set $x_{\nu_0,a,k}=\pm 1$ and $\xi_{\nu,a,k}=1$. This is however not the simplest choice for $\eta$. It is not difficult to check that the following simpler choice works as well.
	\be
	x_{\nu_0,a,k}=\left\{\begin{array}{ccc}
	\pm 1&{\rm for}& k=p_{\nu_0,a}+1\\
	0&&{\rm otherwise}\end{array}\right.,~~~~~
	\xi_{\nu,a,k}=\left\{\begin{array}{ccc}
	1&{\rm for}& k=p_{\nu,a}+1\\
	0&&{\rm otherwise}\end{array}\right..
	\label{e7}
	\end{equation}
In this way one obtains the following set of simple canonical automorphisms with respect to which $H$ is $\eta$-pseudo-Hermitian.
	\bea
	\eta(\sigma)&:=&\sum_{\nu_0}\sum_{a=1}^{d_{\nu_0}}\sigma_{\nu_0,a}\left(
	\sum_{i=1}^{p_{\nu_0,a}} 
	|\phi_{\nu_0},a,i\kt\br\phi_{\nu_0},a,p_{\nu_0,a}+1-i|\right)+\nn\\
	&&\sum_{\nu}\sum_{a=1}^{d_{\nu}}
	\sum_{i=1}^{p_{\nu,a}}
	(|\phi_{\nu},a,i\kt\br\phi_{\nu-},a,p_{\nu,a}+1-i|+
	|\phi_{\nu-},a,p_{\nu,a}+1-i\kt\br\phi_{\nu},a,i|),
	\label{e8}
	\eea
with $\sigma:=\{\sigma_{\nu_0,a}\}$ being a sequence of signs. A straightforward calculation shows that
	\bea
	\eta(\sigma)^{-1}&:=&\sum_{\nu_0}\sum_{a=1}^{d_{\nu_0}}\sigma_{\nu_0,a}\left(
	\sum_{i=1}^{p_{\nu_0,a}} 
	|\psi_{\nu_0},a,i\kt\br\psi_{\nu_0},a,p_{\nu_0,a}+1-i|\right)+\nn\\
	&&\sum_{\nu}\sum_{a=1}^{d_{\nu}}
	\sum_{i=1}^{p_{\nu,a}}
	(|\psi_{\nu},a,i\kt\br\psi_{\nu-},a,p_{\nu,a}+1-i|+
	|\psi_{\nu-},a,p_{\nu,a}+1-i\kt\br\psi_{\nu},a,i|).
	\label{e9}
	\eea

If $H$ is diagonalizable, $p_{n,a}=1$ and (\ref{e8}) yields the expression for the canonical automorphisms given in Ref.~\cite{p4}. Again choosing all the signs $\sigma_{\nu_0,a}$ to be positive yields a positive-semidefinite (nonnegative) $\eta$ and a positive-semidefinite inner product,
	\be
	\bbr\psi,\phi\kkt_\eta:=\br\psi|\eta|\phi\kt.
	\label{e10}
	\end{equation}
However, even if the complex eigenvalues are absent this choice does not lead to a positive-definite inner product on the Hilbert space unless $H$ is diagonalizable. This is because in general there are
defective (real) eigenvalues $E_{\nu_0}$; at least one of the Jordan dimensions $p_{\nu_0,a}$ is greater than 1; and according to (\ref{e5.1}) and (\ref{e5.2}), $\bbr\psi_{\nu_0},a,1|\psi_{\nu_0},a,1\kkt_\eta=\br\psi_{\nu_0},a,1|\eta|\psi_{\nu_0},a,1\kt=0$. Hence the corresponding eigenvector $|\psi_{\nu_0},a,1\kt$ is null, and the  inner product~(\ref{e10}) is not positive-definite.

\begin{itemize}
\item[] {\bf Theorem~2:} Let $H$ be as in Theorem~1. Then $H$ is pseudo-Hermitian if and only if it is Hermitian with respect to a positive-semidefinite inner product $\bbr~,~\kkt:{\cal H}^2\to\C$, i.e., for all $\phi,\psi\in{\cal H}$, $\bbr\phi,H\psi\kkt=\bbr H\phi,\psi\kkt$.
\item[] {\bf Proof:} Suppose $H$ is pseudo-Hermitian, then according to Theorem~1 it has real and/or complex-conjugate pairs of eigenvalues with identical geometric multiplicity and Jordan dimensions. According to the proof of this theorem, this implies that $H$ is pseudo-Hermitian with respect to the automorphism~(\ref{e8}) with $\sigma_{\nu_0,a}=1$ for all $\nu_0$ and $a\in\{1,2,\cdots,d_{\nu_0}\}$. The latter yields the positive-semidefinite inner product~(\ref{e10}) which satisfies, for all $\psi,\phi\in{\cal H}$,
	\[ \bbr\phi,H\psi\kkt_\eta=\br\phi|\eta H|\psi\kt=
	\br\phi|H^\dagger\eta|\psi\kt=\br H\phi|\eta|\psi\kt=\bbr H\phi,\psi\kkt_\eta.\]
Hence $H$ is Hermitian with respect to the  inner product~(\ref{e10}). Conversely, let $H$ be 
Hermitian with respect to a positive-semidefinite inner product $\bbr~,~\kkt$. Let $\eta:{\cal H}\to{\cal H}$ be
defined in terms of its matrix elements according to, for all $\psi,\phi\in{\cal H}$,
	\[\br\psi|\eta|\phi\kt:=\bbr\psi,\phi\kkt.\]
Then, because $\bbr~,~\kkt$ is a sesquilinear, Hermitian, nondegenerate quadratic form \cite{kato}, $\eta$ is a linear, Hermitian, automorphism. Furthermore, because $H$ is Hermitian with respect to $\bbr~,~\kkt$
we have, for all $\psi,\phi\in{\cal H}$,
	\[ \br\phi|\eta H\psi\kt=\bbr\phi,H\psi\kkt=\bbr H\phi,\psi\kkt=\br H\phi|\eta|\psi\kt=
	\br\phi|H^\dagger\eta\psi\kt.\]
Therefore, $\eta H=H^\dagger\eta$ or $H^\sharp:=\eta^{-1}H^\dagger\eta=H$, i.e., $H$ is pseudo-Hermitian.~~~$\square$
\end{itemize}

\section{$2\times 2$ Matrix Hamiltonians}

In Ref.~\cite{p3}, we showed that the pseudo-Hermiticity of a diagonalizable Hamiltonian is equivalent to the presence of antilinear symmetries. The PT-symmetry studied in the literature \cite{pt} is a primary example. In general, such a Hamiltonian depends on certain continuous parameters whose variation does not destroy the symmetry but changes the spectrum. In particular, it is possible that under such variations complex-conjugate pairs of eigenvalues merge and produce real eigenvalues or a real eigenvalue splits into a complex-conjugate pair of eigenvalues. This is a generic behavior observed in the numerical studies of PT-symmetric Hamiltonians \cite{pt} and naturally applies in the case of general pseudo-Hermitian Hamiltonians. Now consider a diagonalizable pseudo-Hermitian Hamiltonian with a discrete spectrum that undergoes a continuous pseudo-Hermiticity-preserving perturbation. In general, such a perturbation may not preserve the diagonalizability of the Hamiltonian \cite{mz}. In particular, at the critical values of the perturbation parameter when two nondegenerate complex-conjugate eigenvalues merge to produce a real eigenvalue, there is no guarantee that the resulting eigenvalue is doubly degenerate. This observation underlies the importance of the results of Section~3 in the study of the behavior of diagonalizable pseudo-Hermitian operators undergoing arbitrary pseudo-Hermiticity-preserving perturbations. 

Consider the case that under such a perturbation a pair of complex-conjugate nondegenerate eigenvalues cross while no other level-crossing occurs. In the vicinity of this level-crossing, one can approximate the behavior of the Hamiltonian by a traceless $2\times 2$ matrix Hamiltonian. In Ref.~\cite{p4}, we have studied the properties of general complex, traceless, diagonalizable, pseudo-Hermitian $2\times 2$ matrix Hamiltonians. A traceless $2\times 2$ matrix $H$ with two nondegenerate eigenvalues is pseudo-Hermitian if its determinant is a nonzero real number, \cite{p4}. As we explain below the converse of this statement is also true. In particular, $\det(H)<0$ or $\det(H)>0$ depending on whether the eigenvalues are real or imaginary. This means that the moduli space ${\cal M}$ of 
traceless pseudo-Hermitian $2\times 2$ matrices with two nondegenerate eigenvalues is a 5-dimensional subspace of the 8-dimensional space $M(2,\C)$ of all complex $2\times 2$ matrices. The latter has the manifold structure of $\C^4=\R^8$. If we respectively denote  the subsets of complex traceless $2\times 2$ matrices, complex traceless pseudo-Hermitian $2\times 2$ matrices, and traceless Hermitian $2\times 2$ matrices by $M_0$, ${\cal M}'$ and ${\cal M}_0$, we have
	\[\begin{array}{ccccccccc}
	{\cal M}_0&\subset &{\cal M}&\subset& {\cal M}'&\subset &M_0&\subset &M(2,\C)\\
		|| &&&&&&||&&|| \\
	\R^3&&&&&&\R^6&&\R^8.
	\end{array}\]

We can identify ${\cal M}$ with the inverse image of $\R-\{0\}\in\R^2=\C$ under the continuous function $\det:M(2,\C)\to\C=\R^2$. Noting that $\R^+$ and $\R^-$ are disjoint, open, connected subsets of $\C$ and $\det$ is continuous, we infer that ${\cal M}$ consists of two open connected components, namely
	\[ {\cal M}^\pm:=\{ H\in {\cal M}~|~\det H\in\R^\pm\}.\]
This in turn implies that at a critical point of the parameters of $H$ where a level-crossing happens, $H$ fails to stay in ${\cal M}$. This is also easily seen by realizing that because $H$ is traceless, a level-crossing can occur only if $\det H$ vanishes. Therefore, at the level-crossing either $H=0$ or it
is nondiagonalizable.

In fact, it is not difficult to see that an element $X_\pm$ of ${\cal M}^\pm$ has the general form
	\be
	X_\pm= \sqrt{\pm 1}~E g^{-1}\sigma_3 g,
	\label{e21}
	\end{equation}
where $E$ is a nonzero real number, $g$ is an element of the special linear group $SL(2,\C)$, and $\sigma_3$ is the diagonal Pauli matrix ${\rm diag}(1,-1)$. The form~(\ref{e21}) indicates that the
moduli spaces ${\cal M}^+$ and ${\cal M}^-$ have the manifold structure of $F\times(\R-\{0\})$ where $F$ is the 4-dimensional homogeneous space
	\[F:=SL(2,\C)/U_{\C}(1),\]
and 
	\[U_{\C}(1):=\left\{e^{z \sigma_3}|z\in\C\right\}=
	\left\{\left.\left(\begin{array}{cc}	
	w & 0\\
	0 & w^{-1}\end{array}\right)~\right| w\in \C-\{0\}\right\}.\]
Furthermore, according to (\ref{e21}) the group elements $g$ that are uniquely parameterized by the points of $F$ play the same role for both $X_+$ and $X_-$. It is the factor $\sqrt{\pm 1}~E$ in (\ref{e21}) that differentiates $X_+$ and $X_-$. This suggests that we can identify  ${\cal M}^\pm$ by $F\times L^\pm$, where 
	\bea
	L^+ &:=&\{ z\in\C-\{0\}~|~ {\rm Re}(z)=0\}=\mbox{imaginary axis in the complex plane with 	$0$ removed},\nn\\
	L^-&:=&\{ z\in\C-\{0\}~|~ {\rm Im}(z)=0\}=\mbox{real axis in the complex plane with $0$ 	removed},\nn
	\eea
and `Re' and `Im' stand for the `real' and the `imaginary' part of the corresponding complex variable, respectively.

The above picture of ${\cal M}$ confirms our earlier remark that at a level-crossing a traceless pseudo-Hermitian $2\times 2$ matrix, 
	\be
	H=\left(\begin{array}{cc}
	a & b\\
	c & -a\end{array}\right), 
	\label{2by2}
	\end{equation}
either vanishes identically:
	\be
	a=b=c=0,
	\label{e22a}
	\end{equation}
or becomes nondiagonalizable:
	\be
	a=\pm i\sqrt{bc},~~~~~|a|^2+|b|^2+|c|^2\neq 0,
	\label{e22}
	\end{equation}
where $i:=\sqrt{-1}$. In the latter case, according to (\ref{H=H_b}), $H$ has the form $H=|\psi_1\kt\br\phi_2|$ where $\{|\psi_a\kt,|\phi_a\kt\}$ with $a=\{1,2\}$ is a complete
biorthonormal system in $\C^2$. In particular, we have
	\begin{itemize}
	\item[] {\bf Proposition~1:} Every traceless nondiagonalizable $2\times 2$ matrix $H$ is 
	pseudo-Hermitian.
	\item[] {\bf Proof:} Because $H$ is both traceless and nondiagonalizable, zero is the only 	eigenvalue of $H$. Hence according to Theorem~1, it must be pseudo-Hermitian.~~~$\square$
	\item[] {\bf Theorem~3} A traceless $2\times 2$ matrix $H$ is pseudo-Hermitian if and only if it 	has a real determinant, i.e.,
		\[{\cal M}'=\left\{H\in M_0~|~\det H\in\R\right\}.\]
	\item[] {\bf Proof:} If $H$ is not diagonalizable, then according to Proposition~1 it is 
	pseudo-Hermitian, and the statement of Theorem~3 is trivially satisfied. If 
	$H$ is diagonalizable, it is either identically zero, in which case it is pseudo-Hermitian and has a 
	real (zero) determinant, or it has two nondegenerate eigenvalues. In the latter case, in view of a 
	proposition proven in Ref.~\cite{p4}, reality of the determinant of $H$ implies its pseudo-	Hermiticity. The converse is also true. For if $H$ is pseudo-Hermitian, then its eigenvalues
	are either both real or they are complex-conjugate of one another. Because $H$ has a vanishing 	trace, in the latter case the eigenvalues must be imaginary. This in turn implies that in both cases 	the determinant of $H$ is real.~~~$\square$
	\end{itemize}

In light of Theorem~3, the possibility~(\ref{e22a}) that at a level-crossing a traceless pseudo-Hermitian $2\times 2$ matrix Hamiltonian remains diagonalizable corresponds to a single point in the uncountably infinite set of traceless nondiagonalizable pseudo-Hermitian $2\times 2$ matrices ${\cal M}'-{\cal M}$. To make this observation more transparent, consider the pseudo-Hermitian matrix Hamiltonians~(\ref{2by2}) corresponding to the choice $c=0$. 
Then $\det H=-a^2$, $a\in L^\pm$, and $H\in{\cal M}^\pm$. Now suppose that $a$ and $b$ are analytic functions of a real perturbation parameter $\lambda$ and that a level-crossing occurs at $\lambda=0$. Then at the vicinity of the level-crossing, i.e., for  $|\lambda|<\epsilon$ for some sufficiently small $\epsilon\in\R^+$,
	\[ a(\lambda)\approx\left\{\begin{array}{ccc}
	a_{\rm r}\lambda&{\rm for} & -\epsilon<\lambda\leq 0\\
	ia_{\rm i}\lambda & {\rm for} & 0\leq\lambda<\epsilon
	\end{array}\right.,~~~~~~b(\lambda)\approx b_0+b_1\lambda,\]
where $a_{\rm r}$ and $a_{\rm i}$ are nonzero real constants, and $b_0$ and $b_1$ are complex constants. At $\lambda=0$, $H$ vanishes identically provided that $b_0=b(0)=0$. This is the only way in which $H$ can maintain its diagonalizability. Clearly, for $b_0\neq 0$, $H$ becomes nondiagonalizable at $\lambda=0$. In both cases $a(0)=0\in\R$ is the only eigenvalue (alternatively $\det H=0$). Hence according to Corollary~1 (respectively Theorem~3), $H$  remains pseudo-Hermitian at $\lambda=0$. This example clearly shows that the loss of diagonalizability at the crossing of the complex-conjugate eigenvalues is a generic behavior.

\section{Application}

Consider the Wheeler-DeWitt equation for the closed FRW minisuperspace model with a real massive scalar field,
	\be
	\left[-\frac{\partial^2}{\partial\alpha^2}+\frac{\partial^2}{\partial\varphi^2}+e^{4\alpha}-m^2\,
	e^{6\alpha}\varphi^2\right]\,\psi(\alpha,\varphi)=0,
	\label{wdw}
	\end{equation}
where $\alpha:=\ln {\rm a}$, `a' is the scale factor, $\varphi$ is a real scalar field of mass $m$, and we have chosen a particularly simple factor ordering and the natural units, \cite{page,wiltshire}. The Wheeler-DeWitt equation~(\ref{wdw}) can be written in the Schr\"odinger form $i\dot\Psi=H\Psi$ where $\Psi$ is the two-component wave function  \cite{jmp-98}
	\[\Psi:=\frac{1}{\sqrt 2}\,\left(\begin{array}{c}
	\psi+i\dot\psi\\
	\psi-i\dot\psi\end{array}\right),\]
$H$ is the effective Hamiltonian
	\be
	H:=\frac{1}{2}\left(\begin{array}{cc}	
	1+D & -1+D\\
	1-D & -1-D\end{array}\right),
	\label{H-wdw}
	\end{equation}
a dot means a derivative with respect to $\alpha$, and 
	\be
	D:=-\frac{\partial^2}{\partial\varphi^2}+m^2\, e^{6\alpha}\varphi^2-e^{4\alpha}.
	\label{D=}
	\end{equation}

The eigenvalue problem for the Hamiltonian~(\ref{H-wdw}) may be easily solved \cite{jmp-98}. The eigenvectors $\Psi_{n\pm}$ and the corresponding eigenvalues $E_{n\pm n}$ have the form
	\be
	\Psi_{n\pm}=\frac{1}{\sqrt 2}\,\left(\begin{array}{c}
	1+E_{n\pm}\\
	1-E_{n\pm}\end{array}\right)\phi_{n},~~~~~~
	E_{n\pm}=\pm\sqrt{m\,e^{3\alpha}(2n+1)-e^{4\alpha}}=
	\pm {\rm a}\sqrt{{\rm a}[m(2n+1)-{\rm a}]},
	\label{eg-value2}
	\end{equation}
where $n=0,1,2,\cdots$, $\phi_{n}:=N_n H_n(m^{1/2}e^{3\alpha/2}\varphi)\,
e^{-m\,e^{3\alpha}\varphi^2/2}$, $H_n$ are Hermite polynomials, and $N_n:=[m\,e^{3\alpha}/(\pi 2^{2n}{n!}^2)]^{1/4}$ are normalization constants. 

As seen from~(\ref{eg-value2}), for ${\rm a}\leq m$ the spectrum of $H$ is real, and for ${\rm a}>m$ it consists of real and complex-conjugate pairs of eigenvalues. In general $H$ is pseudo-Hermitian, because $H^\dagger=\sigma_3 H\sigma_3$. For ${\rm a}\neq (2n+1)m$, it is also diagonalizable. But at the critical values ${\rm a}=(2n+1)m$ where a real (namely the zero) eigenvalue splits into a complex-conjugate pair of eigenvalues or the converse happens, $H$ fails to be diagonalizable. The situation is precisely like the one discussed in Section~4. Here the  perturbation parameter has the form $\lambda:={\rm a}-(2n+1)m$. At the vicinity of a level-crossing where $\lambda\to 0$, the operator $D$ and its eigenvectors do not undergo any discontinuous changes. Therefore, one can approximate the span of the eigenvectors $\Psi_{n-}$ and $\Psi_{n+}$ for $\lambda\neq 0$ with the span of the vectors
	\[ |1\kt=\left(\begin{array}{c}
	\phi_n\\ 0 \end{array}\right),~~~~
	|2\kt=\left(\begin{array}{c}
	0 \\ \phi_n \end{array}\right),\]	
where $\phi_n$ is evaluated at $\alpha=\ln{\rm a}=\ln[(2n+1)m]$, i.e., $\lambda=0$. Clearly, we can study the level-crossing by confining our attention to this subspace. The above approximation becomes exact in the limit $\lambda\to 0$. In the subspace spanned by $|1\kt$ and $|2\kt$ the operator $D$ is identically zero. Therefore, we can approximate $D$ by a constant that tends to zero as $\lambda\to 0$. Therefore, the Hamiltonian~(\ref{H-wdw}) takes the form of the matrix Hamiltonian~(\ref{2by2}) with $a=(1+D)/2,  b=(-1+D)/2$,  and $c=(1-D)/2$. In the limit $\lambda\to 0$, $D$ approaches zero, and the conditions~(\ref{e22}) hold. Hence, as expected, $H$ becomes nondiagonalizable at the level-crossing. 

The above argument implies that in general $H$ is diagonalizable for all values of the scale factor except the critical values ${\rm a}= (2n+1)m$. At these values $H$ becomes nondiagonalizable as one of its eigenvalues, namely the zero eigenvalue,  becomes defective. The algebraic multiplicity of this eigenvalue is two. In fact, the effective Hamiltonian~(\ref{H-wdw}) belongs to the class of block-diagonalizable Hamiltonians discussed in Sections~2 and 3. Its canonical Jordan form consists of a $2\times 2$ Jordan block corresponding to the zero eigenvalue and an infinite number of trivial ($1\times 1$) blocks corresponding to nonzero eigenvalues. The fact that this Hamiltonian is pseudo-Hermitian for all values of the scale factor, its spectrum consists of real and complex-conjugate eigenvalues, and its complex eigenvalues are not defective is consistent with the general results of Section~3.

\section{Summary and Conclusion}

In this article we generalized our earlier results on diagonalizable pseudo-Hermitian Hamiltonians to a broad class of nondiagonalizable Hamiltonians. We showed that if a pseudo-Hermitian Hamiltonian may be mapped to a block-diagonal operator with finite-dimensional blocks via a similarity transformation, then the characterization theorems of Ref.~\cite{p1} apply provided that the number and size of the Jordan blocks for the complex-conjugate pairs of eigenvalues are identical.

We also discussed the implications of our findings for the phenomenon of the loss of diagonalizability at the crossing of the complex-conjugate pairs of eigenvalues of diagonalizable pseudo-Hermitian Hamiltonians. For the latter pseudo-Hermiticity is known to be equivalent to the presence of an antilinear symmetry \cite{p3}. This in particular means that our results are relevant in the description of the $PT$-symmetric systems that are diagonalizable except in case of level-crossings of the complex-conjugate eigenvalues due to perturbations of the Hamiltonian. If at the critical values of the perturbation parameter each level-crossing involves a finite number of levels, then our results apply generally. This seems to be the case for various $PT$-symmetric models studied in the literature.
Specifically, at the critical values of the parameters of the $PT$-symmetric systems that undergo a spontaneous $PT$-symmetry breaking, a pair of real eigenvalues merge and a loss of diagonalizability similar to the one discussed in Section~5 occurs.

As a final note, we wish to emphasize that the results of this paper rely on the basic assumption that the quantum system has a genuine separable Hilbert space in which the Hamiltonian acts. For many $PT$-symmetric Hamiltonians the (inner product) structure of the function space in which one solves for the eigenfunctions is not clear. In this context, the assumption of considering non-Hermitian Hamiltonians acting in a separable Hilbert space may seem too restrictive. Nevertheless, we believe that this assumption provides a framework for exploring some of the intriguing properties of a class of non-Hermitian Hamiltonians. This class includes many $PT$-symmetric Hamiltonians as well as all the matrix Hamiltonians and the non-Hermitian Hamiltonians appearing in the two-component formulation of the Klein-Gordon and Wheeler-DeWitt equations. 

\section*{Acknowledgment}
I wish to thank the anonymous referee(s) of  Refs.~\cite{p3} and \cite{p4}, who emphasized that the
diagonalizability assumption made in these papers might be a serious limitation, and M.~Znojil who convinced me that a loss of diagonalizability occured in certain $PT$-symmetric models. Finally I would like to acknowledge the support of the Turkish Academy of Sciences through the Young Researcher Award Program (GEBIP).

\ed